\documentclass[aps,twocolumn,notitlepage,superscriptaddress,nogroupedaddress,nofootinbib]{revtex4-2}

\usepackage{xcolor,graphicx}
\usepackage{amssymb,amsmath,graphicx}
\usepackage[utf8]{inputenc}
\usepackage{comment}
\usepackage{braket}
\usepackage[normalem]{ulem}
\usepackage{hyperref,url}
\usepackage[bb=dsserif]{mathalpha}

\usepackage{tikz}
\usetikzlibrary{positioning,arrows.meta,bending}

\usepackage[caption = false]{subfig}

\usepackage{cleveref}
\usepackage{physics}

\begin{document}

\title{Enhancing Quantum Annealing via entanglement distribution}

\author{Raúl Santos}
\affiliation{Instituto Superior Técnico, Universidade de Lisboa, Portugal}
\affiliation{Physics of Information and Quantum Technologies Group, Centro de Física e Engenharia de Materiais Avançados (CeFEMA), Portugal}
\affiliation{PQI -- Portuguese Quantum Institute, Portugal}
\affiliation{Technical University of Eindhoven, Netherlands}

\author{Lorenzo Buffoni}
\affiliation{PQI -- Portuguese Quantum Institute, Portugal}
\affiliation{Department of Physics and Astronomy, University of Florence, 50019 Sesto Fiorentino, Italy}

\author{Yasser Omar}
\affiliation{Instituto Superior Técnico, Universidade de Lisboa, Portugal}
\affiliation{Physics of Information and Quantum Technologies Group, Centro de Física e Engenharia de Materiais Avançados (CeFEMA), Portugal}
\affiliation{PQI -- Portuguese Quantum Institute, Portugal}

\begin{abstract}

Quantum Annealing has proven to be a powerful tool to tackle several optimization problems. However, its performance is severely impacted by the limited connectivity of the underlying quantum hardware, compromising the quantum speedup. In this work, we present a novel approach to address these issues, by describing a method to implement non-local couplings throught the lens of Local Operations and Classical Communcations (LOCC). Non-local couplings are very versatile, harnessing the configurability of distributed quantum networks, which in turn lead to great enhancement of the physical connectivity of the underlying hardware. Furthermore, the realization of non-local couplings between distinct quantum annealing processors activates the scalability potential of distributed systems, i.e. allowing for a distributed quantum annealing system. Finally, in a more distant vision, we also show that secure multi-party quantum annealing algorithms are possible, allowing for cooperation of distrusting parties through optimization with quantum annealing and a particular type of non-local couplings.

\end{abstract}

\maketitle

\definecolor{mygray}{gray}{0.3}

\section{Introduction}


Solving optimization problems is a very common problem in science with many practical applications \cite{korte2011combinatorial,paschos2014applications}, field to which physics has contributed with lots of concepts and methods, starting from the idea of thermal simulated annealing \cite{Kirkpatrick671}, to the applications of replica and cavity methods \cite{Krza13}. Quantum Annealing was initially formulated in the 90's \cite{Nishi98} as a quantum alternative to classical simulated annealing in which quantum tunneling replaces thermal hopping in order for the system to avoid being trapped in local minima and reach the ground state (i.e. the solution of the optimization problem). This idea was later developed to the point at which, today, we have working quantum annealers with thousands of qubits available.

However, Quantum Annealers suffer from some issues that  are limiting the size and complexity of the problems we are currently able to solve efficiently using this technology.  The  major problems that we will be addressing here are embedding and connectivity, which are closely related. The connectivity problem is simply an engineering bound to the number of qubits that one is effectively able to couple on the chip layout - in the latest generation quantum annealers the number of couplings per qubit is $\sim 15$. This limits the problems that one can solve on a quantum annealer to ones that are sparsely connected. A way to circumvent this problem is to use minor-embedding techniques in which a chain of physical qubits is treated as one logical qubit allowing for more connections. This technique, however, is pretty intense in the number of qubit used thus limiting the size of the problems which we can effectively solve. Another issue comes from the restrictions originating from the fabrication techniques of superconducting chips: the topology of the coupling graphs is quasi-planar in nature, and that can be an issue in achieving a quantum speedup in the finite-temperature regime \cite{katzgraberGlassyChimerasCould2014}.

Another concern is, as quantum processors increase in size, it becomes harder and harder to conserve the quantum information stored in the qubits throughout the computation, due to noise and decoherence effects. There is a point where increasing the number of qubits in a single processor becomes impractical. Inspired by the successes of classical high performance computing, such as traditional supercomputers, Distributed Quantum Computation (DQC) aims to improve the scalability of a quantum computing technology by employing it in a cluster of quantum processors, interconnected via quantum channels capable of distributing entanglement \cite{kimbleQuantumInternet2008,vanmeterPathScalableDistributed2016,wehnerQuantumInternetVision2018a,cuomoDistributedQuantumComputing2020a,nemotoDistributedQuantumComputation2020}. These, joined by classical communication channels and quantum operations localized on the quantum processors, allow for a long-distance quantum operations scheme called Local Operations and Classical Communications (LOCC) \cite{eisertOptimalLocalImplementation2000,Yang2015,piroli2021}. Parallel quantum operations become possible due to quantum correlations being generated between processors through these LOCC operations. 

In an early stage these distinct quantum processors may make part of a multi-core quantum processing unit (QPU). This technology is already on the horizon, with IBM planning to develop multi-core quantum machines as early as 2024 \cite{ExpandingIBMQuantum2022}. At a later stage, clusters of multi-core quantum computers would be available in a similar fashion as traditional supercomputers in a Quantum Local Area Network (QLAN). In a more futuristic vision, as the quantum internet becomes more widespread \cite{kimbleQuantumInternet2008, wehnerQuantumInternetVision2018a,Singh2021QuantumDirections,illianoQuantumInternetProtocol2022,wangApplicationScenariosQuantum2023}, with applications such as elections \cite{Tani2005ExactProblem}, quantum key distribution \cite{Bennett2014QuantumTossing,Sharma2019ADistribution} and any other that may require a wide-spread quantum network to transfer quantum data, the DQC scheme could grant novel secure multi-party applications, where two or more parties want to collaborate by solving a common optimization problem without sharing secret, securely kept, data, between the distrusting parties \cite{wenliangduPrivacypreservingCooperativeScientific2001}.


In this work, we show that through the capability of non-local couplings, all of these promises of DQC can also be applied to quantum annealing processes. This non-local coupling capability may also unlock the quantum speedup that quantum annealers have been having difficulty in achieving. After a brief introduction to quantum annealing and distributed quantum computation in \cref{sect:background}, we show, in \cref{sect:methods} how these non-local couplings can be implemented through LOCC. \cref{sect:applications} discusses potential applications and their advantages. Mathematical derivations, numerical simulations and examples of secure multi-party problems for quantum annealing are shown in the Appendix.

\section{Background} \label{sect:background}


\subsection{Quantum Annealing}

Quantum annealers are a particular kind of analog quantum computers, where their inherent process is continuous in time, governed by a time-dependent slow varying (adiabatic) Hamiltonian $H(t)$. 
\begin{equation}
    H(t) = \left( 1 - \frac{t}{t_F}\right) H_0 + \frac{t}{t_F} H_F. \label{eq:annealing_Ht}
\end{equation}

 The system of qubits is modeled with the transverse field Ising Model. It consists of an initial Hamiltonian $H_0$, often referred to as the mixing Hamiltonian, and its ground state is the starting point of the annealing evolution. There is also a final Hamiltonian $H_F$, called target or objective Hamiltonian. Quantum annealing corresponds, in essence, to a quantum search, where the target is the ground state of the final Hamiltonian $H_F$. The mixing Hamiltonian $H_0$ dictates the search space of the annealing process. In the standard implementations of quantum annealing, $H_0$ and $H_F$ take the form of \cref{eq:annealing_H0,eq:annealing_HF}, respectively. The $h_i$ terms correspond to local fields applied to the qubit $i$. With 2-body Hamiltonians, we have the couplings $J_{ij}$ which quantify the strength of the $\sigma^z\sigma^z$ interactions between the qubits $i,j$. We label the qubits with the set $\Omega=\{0, 1, ...\}$. These are connected by the set of edges $e$ that indicate which qubits are coupled, i.e. $e=\{ (i,j): J_{ij}\ne 0\}$.

\begin{equation}
    H_0 = \sum_{i\in\Omega} \sigma_i^x \label{eq:annealing_H0}
\end{equation}
\begin{equation}
    H_F = \sum_{i\in\Omega} h_i \sigma_i^z + \sum_{(i,j)\in e} J_{ij} \sigma_i^z \sigma_j^z \label{eq:annealing_HF}
\end{equation}
  

In the context of computation, this process is useful as an optimization meta-heuristic, where the final Hamiltonian $H_F$ encodes an objective function. Furthermore, having $H_F$ expressed in the Ising Model, quantum annealing becomes an ideal process for solving Quadratic Unconstrained Binary Optimization (QUBO) problems \cite{gloverQuantumBridgeAnalytics,lucasIsingFormulationsMany2014}. These problems, other than being generally hard to solve, are often defined on graphs with a large number of connections. This limits the use of current generation quantum annealers as they suffer from connectivity problems as discussed above.


There exist a few platforms which are capable of implementing quantum annealing processes. The superconducting platform, lead by DWave \cite{dwaveSalesman2021,dwaveBenchmark2022}, has achieved on the order of thousands of qubits on a single chip, due to lithographic techniques. In order to keep the coherence of the qubits, these have to operate at cryogenic temperatures. The manufacturing process limits the topology of the connectivity of the qubits to quasi-planar graphs, leading to the aforementioned issues regarding limited qubit connectivity. The cold atoms and trapped ions platforms are also capable of quantum annealing \cite{glaetzleCoherentQuantumAnnealer2017,torgglerQuantumNQueensSolver2019}, although at a smaller scale as of yet (tens to hundreds of qubits). They are placed in an arbitrary lattice by the use of electromagnetic fields, be they microwave, laser light (cold atoms) or electric fields generated by electrodes (trapped ions). These platforms are very versatile, but are still on their infancy, and it is not clear how they compare against the superconducting platform in the long term. Although, one important advantage for the protocol we propose in this article is the capability to communicate non-locally through photons. The following section discusses more on this capability.  



\subsection{Distributed Quantum Computation}

A network of interconnected quantum nodes, communicating through quantum channels is capable of novel applications which are not available through classical networks (like the classical internet). Namely, the quantum internet will allow for quantum cryptography applications  and novel quantum computing applications not possible through classical channels \cite{wangApplicationScenariosQuantum2023}.  Distributed quantum computation allows for the use of smaller quantum computers with more manageable noise and in principle could then obtain better qubit coherence.  Arbitrary quantum operations between two distant parties can be achieved by having a classical and a quantum channel between both parties \cite{eisertOptimalLocalImplementation2000}. Aiming to tackle the scalability problem of their quantum computers, IBM plans in their roadmap for 2024, to introduce the 462-qubit processor architecture "Flamingo", supporting quantum communication links 
\cite{ExpandingIBMQuantum2022,IBMQuantumComputing}.

\begin{figure}[t]
    \centering
    \includegraphics[width=\linewidth]{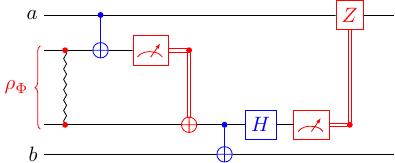}
    \caption{CNOT distribution protocol proposed in \cite{eisertOptimalLocalImplementation2000}. In blue are the local operations, realized locally in each quantum computer, while in red are all the elements necessary for the distribution itself: shared entanglement, measurement and classical communication.}
    \label{fig:DCNOT-circuit}
\end{figure}

In the quantum internet quantum communication is realized through photon links that share the entanglement resource, allowing for arbitrary quantum links to be made between any set of nodes in the network, through quantum operations and measurement \cite{Caleffi2017OptimalNetworks,Hahn2019QuantumComplementation,Pant2019RoutingInternet}. These entangled links are then used to implement the desired quantum protocols (eg. quantum teleportation, distributed CNOT). The class of protocols which execute distributed operations through quantum networks, local operations and measurement are called Local Operation and Classical Communication (LOCC), which we will consider for the long distance interactions in our work. An arbitrary quantum circuit can be executed exclusively through single qubit unitary rotations and two two-qubit entangling gates, such as CNOT \cite{tucciIntroductionCartanKAK2005}. Thus, the problem of distributing an arbitrary circuit can be reduced to CNOT distribution. Reference \cite{eisertOptimalLocalImplementation2000} describes just how that would be possible through the LOCC framework, in a protocol which is shown in \cref{fig:DCNOT-circuit}. This operation is equivalent to a CNOT gate when the fidelity of the shared entangled pair is $F_\Phi=1.0$. Distribution is sensitive to the noise of the generated entangled pair, the latter poses as one of the most relevant factor on the feasibility and practicality of the applications for Distributed Quantum Computing and the Quantum Internet in general. Another of the main challenges of distributed quantum computation is to have an efficient and low noise interface between the entanglement carriers and the computing qubits.

There are a few platforms available that have shown to be capable of computation as well as long distance communication. Firstly there are the cold atoms and trapped ions platforms, which are able to communicate non-locally through Cavity QED methods \cite{reisererCavitybasedQuantumNetworks2015,weltePhotonMediatedQuantumGate2018}. For near term applications, where distribution may occur within a single processor as a way to enhance connectivity, cavity QED and atom/ion shuttling could be viable methods to achieve this in the monolithic processor regime. 

For the superconducting platform, the scenario is different. A possible approach to connect distant superconducting qubits is by having a superconducting wave-guide cable connecting them \cite{majerCouplingSuperconductingQubits2007,magnardMicrowaveQuantumLink2020}. This technology is not very practical for long distance communication, but could be useful in multi-core setups or even in Quantum LANs. Another interesting approach is by the use of opto-mechanical transducers, which couple communication photons with the solid-state superconducting qubits used in the computations \cite{stannigelOptomechanicalTransducersQuantum2011,singhOptomechanicalCouplingMultilayer2014,fiaschiOptomechanicalQuantumTeleportation2021}. Due to the use of photons as the quantum communication medium, this technology would be more viable for long-distance quantum communication. 


Within the context of distributed quantum computation we will describe how a non-local coupling can be simulated. These non-local couplings inherit the versatility of the underlying distributed entanglement network, allowing for arbitrary connectivity between qubits with a non-local interface (\textit{interface} qubits). It might be the key ingredient to unlock the  theorised speedups of quantum annealing. In the next section we describe the theory behind non-local couplings: how the process works, potential applications and its limitations.    




\section{Distributing non-local couplings} \label{sect:methods}



\begin{figure}
    \centering
    \includegraphics[width=\linewidth]{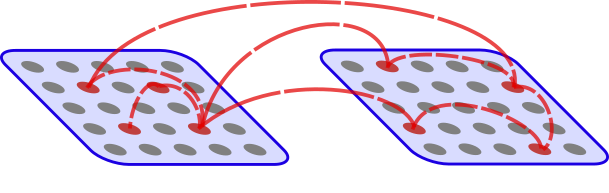}
    \caption{Example of a configuration for a distributed quantum annealing network, with two annealing systems encapsulated in blue regions. In this representation, the qubits are coupled through nearest neighbour interactions, leading to a planar connectivity topology. The red dashed lines with short dashes represent non-local couplings within the same annealing system, as a way to enhance the connectivity of the hardware. The red long-dashed lines represent non-local couplings between qubits in distinct systems. The red points inside the qubit nodes indicate that those qubits are interface nodes, therefore being able to communicate non-locally. The interface qubits can communicate to any other interface qubit reachable by an entanglement distribution network, which immensely increases their connectivity relative to local qubits. }
    \label{fig:example-net}
\end{figure}

Assume that we have a continuously varying time-dependent Hamiltonian $H(t)$, which can be separated into local $H_L(t)$ and non-local $H_N(t)$ components such that $H(t)=H_L(t)+H_N(t)$. The local Hamiltonian describes the evolution through the local fields and couplings which are natively available to the annealing hardware. The non-local Hamiltonian represents couplings that are simulated through distribution (LOCC). It may couple qubits within the same processor, or qubits between separate processors, depending on the application. Fig. \ref{fig:example-net} visualizes the distinction between local and non-local Hamiltonians.


The continuous evolution given by $H(t)$ can be expressed as an unitary operator, with the use of a time-ordered exponential  (\ref{eq:trotter_Ut}). The process can then be digitized into a sequence of $M$ unitary operators (\ref{eq:trotter_Ut_prod}).

\begin{align}
    U(0, t_F) &= \mathcal{T}_t \exp \left\{ -i \int_0^{t_F} dt H(t)  \right\} \label{eq:trotter_Ut} \\
    &= \mathcal{T}_{t_k} \prod_{k=0}^{M-1} U(t_{k}, t_{k+1}) \label{eq:trotter_Ut_prod} 
\end{align}
where $\mathcal{T}_t$ is the time-ordering operator, $t_k= k\Delta t_k$ and $$\sum_{k=0}^{M-1}\Delta t_k = t_F,$$ with $\Delta t_k$ is the Trotter step size. Then, each $U(t_k, t_{k+1})$ is Trotterized into the local and non-local parts

\begin{equation}
    U(t_k, t_{k+1}) \to U_N(t_k, t_{k+1})\ U_L(t_k, t_{k+1}).
\end{equation}
So that we get the whole Trotterized evolution

\begin{align}
    \widetilde{U}(0, t_F) &= \mathcal{T}_{t_k} \prod_{k=0}^{M-1}\ U_N(t_k, t_{k+1})\ U_L(t_k, t_{k+1}). \label{eq:trotter-evol}\\
    U_*(t_k, t_{k+1}) &\simeq \exp \Big\{ -i\ H_*(t_k) \ \Delta t_k  \Big\} \quad *\in\{L, N\} \label{eq:evol-approx}
\end{align}

The approximation in \cref{eq:evol-approx} is valid as long as the considered process is adiabatic for the interval $[t_k, t_k+\Delta t_k]$ (otherwise it would have to be described with a time-ordered exponential, or an higher-order approximation). Since our main application is for quantum annealing, we take the approximation valid. The local unitary evolution operators $U_L(t_k, t_{k+1})$ represent a partial evolution of the quantum state inside the quantum annealing processor through the physics of its hardware. These alternate with the non-local unitary operators $U_N(t_k, t_{k+1})$, which simulate and distribute the non-local couplings between qubits within and between quantum processors. These are implemented via a unitary distribution protocol, i.e. telegates, such as the one proposed in \cite{eisertOptimalLocalImplementation2000}. $U_N$ is a generic two-qubit unitary; it can be decomposed into two CNOT gates interleaved with single qubit gates \cite{tucciIntroductionCartanKAK2005}. Therefore, the distributed annealing process corresponds to the fast switching between local $H_L$ and non-local $H_N$ Hamiltonians, the latter which are implemented via telegates. 

The Trotterized evolution introduces errors due to $H_L$ and $H_N$ not commuting with each other. At each time interval  $t\in[t_k, t_{k+1}]$, there may be an equivalent Hamiltonian $H_{eq}(t_k)$ that governs the process as if it was a continuous adiabatic process (\ref{eq:equiv-Hamiltonian}). 
\begin{align}
    -i\ H_{eq} (t_k) \Delta t_k = \ln \Big[\, U_N(t_k, t_{k+1})\, U_L(t_k, t_{k+1}) \,\Big] \label{eq:equiv-Hamiltonian}
\end{align}
The equivalent Hamiltonian $H_{eq}(t)$ can be described by the Baker-Campbell-Hausdorff series. With uniform time-step $\Delta t_k = \Delta t=t_F/M$, the existence of $H_{eq}(t)$ is guaranteed \cite{thompsonConvergenceProofGoldberg1989} for large switching frequencies (i.e. small step size $\Delta t$; see derivations in \cref{apx:convergence-equiv-Hamiltonian}) 
\begin{equation}
    || H_{eq}(t_k) ||_L < \infty \iff t_{k+1}-t_k = \Delta t < \Delta t_M
\end{equation}
where
\begin{equation}
    \Delta t_M = \min_k \Big(\left|\left|H_L(t_k)\right|\right|_L^{-1}, \left|\left|H_N(t_k) \right|\right|_L^{-1}\Big).
\end{equation}
$||\cdot||_L$ is a Lie norm, satisfying 
\begin{equation}
    \big|\big|\, [X, Y]\, \big|\big|_L = || X\,Y - Y\,X ||_L \leq ||X||_L\ ||Y||_L\
\end{equation}
Furthermore, when $\Delta t/\Delta t_M\ll 1$, we get the bound
\begin{equation}
    \left|\left| H_{eq}(t_k) - H(t_k) \right|\right|_L \leq \frac{\Delta t}{(\Delta t_M)^2}\quad \forall\, 0\leq k\leq M. \label{eq:methods:Hdiff}
\end{equation}
This indicates that the equivalent Hamiltonian gets arbitrarily close to the adiabatic Hamiltonian for arbitrarily small $\Delta t$. Thus, in ideal scenarios (no noise and no restrictions for $\Delta t$), the Trotterized process approximates quantum annealing.

In a real-world scenario, the distribution is affected by noise, and so it is useful to understand how robust is the protocol in noisy situations. Each non-local $2$-body coupling, distributed through a non-local unitary, is synthesized into two CNOT telegates and single qubit rotations. The CNOT telegate, as proposed in \cite{eisertOptimalLocalImplementation2000}, consumes a Bell state and two classical bits are exchanged. We consider the noise to come solely from this shared state, which is described with the density operator $\rho_\Phi$

\begin{equation}
    \rho_\Phi = x \ketbra{\Phi}{\Phi} + \frac{1-x}{4}\ \mathbb{I},
\end{equation}
It has a fidelity $F_\Phi= x + \frac{1-x}{4}$ relative to the Bell state ${\ket{\Phi} = (\ket{00}+\ket{11})/\sqrt{2}}$.

Given large enough number of Trotterization steps $M$, such that ideal distribution approximates quantum annealing, the distribution fidelity $p_N(F_\Phi)$ corresponds to the probability of measuring the solution state found by quantum annealing, and has the lower bound (motivated in \cref{apxsect:noise-derivations}) 
\begin{equation}
    \resizebox{0.95\hsize}{!}{$
    p_N(F_\Phi) > x^{\alpha M}  = \left(\,1-\frac{4}{3}(1-F_\Phi)\,\right)^{\alpha M}\sim 1-\frac{4}{3} \alpha M (1-F_\Phi),$}
\end{equation}
where $\alpha$ is the number of consumed entangled pairs required to implement the $U_N(t_k, t_{k+1})$ telegate. Given a certain annealing time $t_F$, the number of Trotter steps becomes $M=t_F/\Delta t_M$. This means that the distributed process requires, on average $\big[p_N(F_\Phi)\big]^{-1}\sim 1+(4/3)\,\alpha\, t_F/\Delta t_M \, (1-F_\Phi)$ runs until the anneal state is observed.  Analytical and numerical results are shown in \cref{apxsect:mathematical,apxsect:numerical} which go into further detail into this bound.

This procedure of switching between a continuous Hamiltonian (e.g. the local $H_L(t)$) and digital gates (e.g. the unitary circuit implementation of $U_N(t_k, t_{k+1})$) has been discussed before in the literature \cite{barendsDigitizedAdiabaticQuantum2016,arrazolaDigitalAnalogQuantumSimulation2016,parra-rodriguezDigitalAnalogQuantumComputation2020, babukhinHybridDigitalanalogSimulation2020,galiciaEnhancedConnectivityQuantum2020, garcia-molinaNoiseDigitalDigitalAnalog2021,yuSuperconductingCircuitArchitecture2022}, being known as Digital-Analog Quantum Computation (DAQC). Reference \cite{parra-rodriguezDigitalAnalogQuantumComputation2020} proposes a model where they keep the continuous Hamiltonian active during the action of the digital gates as, in principle, would be more practical experimentally. References \cite{barendsDigitizedAdiabaticQuantum2016, babukhinHybridDigitalanalogSimulation2020} implements the DAQC protocol experimentally.

The entanglement cost of the protocol is its main limitation for its implementation in the NISQ era. To aleviate this, entanglement efficient distribution via LOCC parallelization could be implemented, following the work in \cite{wuEntanglementefficientBipartitedistributedQuantum2023a}. It may also be possible to reduce the entanglement cost from the following observation: each distribution step $U_N(t_k, t_{k+1})$ is close to the identity (due to small $\Delta t$), so it may be possible to use less entanglement per step by employing weak measurements on the LOCC protocol.

\subsection{Types of non-local couplings}

Any two-qubit unitary can be implemented with the digitally distributed gates. In this section we discuss the two kinds of non-local couplings that can be considered when developing an annealing system with non-local capabilities. There are energy couplings and duplication couplings, and each of these can also take the form of a many-body coupling. 

\subsubsection{Energy coupling}

Takes the form $J \sigma_i^z \sigma_j^z$ and is implemented as a final Hamiltonian (meaning it has a time dependence of $J(t) \propto t/t_F$). This type of coupling is the one usually considered to be available to the hardware. Qubits coupled through a non-local energy coupling evolve similarly to qubits coupled in the hardware. It provides enhanced connectivity for the quantum annealer and can be advantageous even if a small portion of the qubits on the hardware have a non-local interface \cite{katzgraberHowSmallWorldInteractions2018}, potentially unlocking a quantum speedup for quantum annealers. It is more versatile than the hardware couplings, as any interface qubit can be coupled non-locally to any other interface qubit in the processor/cluster. Each non-local coupling has a fundamental cost on the shared entanglement required for the hardware and therefore the overall number of non-local couplings that can be effectively employed is limited if full coherence is required.


\subsubsection{Duplication coupling} \label{sec:duplication-coupling}

Can be described with a joint mixing term $ \sigma^x \sigma^x$, $\sigma^y\sigma^y$, or any other that has the Bell state $\ket{\Phi^\pm}=(\ket{00}\pm\ket{11})/\sqrt{2}$ as a ground state. The $(\pm)$ phase is arbitrary, as the useful property is that the qubits connected by the couplings take the same value. The total Hamiltonian of the system cannot have other mixing terms ($\sigma^x$ or $\sigma^y$) that act on `duplicated' qubits. This is because additional mixing terms induce unwanted transitions that will change the correlated state $\ket{\Phi^\pm}$ into an anti-correlated state $\ket{\Psi^\pm}$ over time. This also means that a qubit A which is already a duplication of a qubit B cannot be duplication-coupled to a qubit C, and so other tricks have to be used to achieve $n$-plication for $n>2$ (namely, the distribution of a $n$-body coupling like $\bigotimes^n\sigma^x$). Another important factor is that the annealing process has to start with the duplicated qubits on the ground state of the joint mixing term (a Bell state or a GHZ state for larger $n$). These conditions guarantee that, in a noiseless setting, the state after evolution will have the duplicated qubits with the same value, as long as the final Hamiltonian only has classical terms (sum of products of  $\sigma^z$). 

In this non-local coupling, each qubit adds more total couplings to the system than in energy couplings. Given a lattice connectivity of $k$, each non-local duplication adds $k$ new neighbours. Therefore, larger connectivities can be achieved with less non-local connections, bringing an overall benefit due to more efficient entanglement resource utilization.



\subsubsection{Many-body couplings}

There can be considered many-body energy couplings $J \sigma^z \cdots \sigma^z$ which often are not available to the native hardware, or multiplication couplings ${\sim \sigma^x\cdots \sigma^x}$. These keep multiple qubits on the state space $\{\ket{0\cdots0}, \ket{1\cdots1}\}$ throughout the evolution, as long as they start on the state $\ket{GHZ^\pm}$ and there are no other mixing terms acting on these qubits. Although, it should be noted that it is not trivial to implement the digital equivalent of the Hamiltonian evolution through these couplings. 

\section{Applications} \label{sect:applications}

\subsection{Short term application: Single QPU}

The first application where the non-local coupling process can be effective is by enhancing the connectivity of interface qubits within a quantum processor, as shown in the short range links of \ref{fig:example-net}. The versatility of the distributed network allows for arbitrary connectivity between interface qubits. The speedup that can be achieved in modern quantum annealers is conditional on whether the physics of the Ising spin-glass encoded by the Hamiltonian allows for a finite-temperature phase transition \cite{katzgraberGlassyChimerasCould2014}. In reference \cite{katzgraberHowSmallWorldInteractions2018}, the authors show that the addition of few non-local couplings on a quasi-planar graph results on a topology where a disordered Ising system can undergo a finite-temperature spin-glass transition, even when an Ising spin-glass does not display a phase transition at any finite temperature for this topology. In other words, the addition of a few non-local couplings between interface qubits within a single processor can potentially unlock a finite-temperature phase transition not available otherwise \cite{katzgraberHowSmallWorldInteractions2018}, leading to a speedup relative to classical methods \cite{katzgraberGlassyChimerasCould2014}. 

\subsection{Intermediate term application: local QPUs}

Eventually, as the number of qubits on a quantum processor increases, it will become harder and harder to deal with unwanted coupling effects between the many qubits embedded in the bulk of the processor. This is especially true for superconducting circuits \cite{Kjaergaard2020SuperconductingPlay}. Akin to the approaches in classical computing, a distributed approach to quantum annealing is possible through the design of non-local couplings. On each QPU there are local couplings, available on the hardware, and interface qubits which are capable of communicating between any other interface qubit on the cluster. The available number of qubits to the annealing system then scales linearly with the number of annealing nodes on the cluster. A non-trivial problem on this application is the question of how to optimally map the original annealing problem, which does not fit on a single annealer node, onto the whole cluster of quantum annealers, considering energy and duplication couplings as possible means of interaction. In this graph splitting problem we can split by edges (energy coupling) or by vertices (duplication coupling) in a way that partitions the original graph into multiple smaller ones. The best graph split should have the largest probability of measuring the annealled state $\left[ p_N(F_\Phi)\right]^{-1}$, or equivalently, the smallest $\alpha/\Delta t_M$, with $\alpha$ the number of consumed entangled pairs per $U_N$ step, and  $\Delta t_M$ the convergence time-step of the equivalent Hamiltonian $H_{eq}$.

\subsection{Long term application: distant QPUs}

In a more distant vision, enabled by a far reaching quantum internet, it becomes possible to connect distant quantum annealers through non-local couplings. An immediate application for this capability is the implementation of secure multi-party quantum annealing, where two or more distant quantum annealers, belonging to distinct distrusting organizations, but willing to cooperate, would be able to jointly solve an optimization problem while keeping sensitive data secret. Secure multi-party quantum annealing is particularly interesting for the realm of cooperative scientific computations \cite{wenliangduPrivacypreservingCooperativeScientific2001}, where two distrusting parties want to collaborate and find an optimal procedure, without revealing the secret information, necessary for the optimization, to the opposing party. These scientific computation problems are often posed as a linear system of equations, which can be solved with the Linear Least Squares approach, easily put into QUBO form, ready for annealing (see \ref{apx:sect:linear-least-squares}). 





\section{Conclusions} 
In this work we have proposed a new protocol that allows for non-local couplings between distant qubits. These may be located within the same annealing processor, but unable to directly couple through the annealing hardware, or in distant quantum processors, leading to different applications of the protocol. We have shown that this can be done through distributed digital operations, such as the distributed CNOT gate, and that the amount of the quantum communication resource required for the protocol scales with the annealing time $t_F$, with the Trotterization time step parameter $\Delta t$ and of course, with the number of non-local couplings that are considered on the system. We envision three major stages of application of this protocol: near term, medium term and long term. In the near term, non-local couplings could be applied between qubits within the same processor in order to improve the dynamics of the Ising spin-glass system. In the medium term, quantum annealing processors interconnected with quantum channels would enhance the scalability of the overall system, due to more manageable noise sources. In fact, IBM plans to take advantage of this point in their quantum roadmap. For the long term vision, with long distance non-local couplings, quantum annealers could execute secure multi-party quantum annealing algorithms, where a quantum annealing speedup would improve the cooperation capabilities between distrusting companies.

\section*{Acknowledgments} 
The authors thank the support from FCT -- Funda\c{c}\~{a}o para a Ci\^{e}ncia e a Tecnologia (Portugal), namely through project UIDB/04540/2020, as well as from projects QuantHEP and HQCC supported by the EU QuantERA ERA-NET Cofund in Quantum Technologies and by FCT (QuantERA/0001/2019 and QuantERA/004/2021, respectively), from the EU Quantum Flagship projects QIA (820445) and QMiCS (820505). This work is also funded by the Advanced Computing/EuroCC MSc Fellows Programme, which is funded by EuroHPC under grant agreement No 951732. L.B. was funded by PNRR MUR Project No. SOE0000098-ThermoQT financed by the European Union--Next Generation EU.


\appendix \label{sect:appendix}

\section{Mathematical derivations}\label{apxsect:mathematical}

\subsection{Convergenge conditions of the equivalent Hamiltonian}\label{apx:convergence-equiv-Hamiltonian}


Since the local and non-local Hamiltonians do not commute, i.e. $[H_L(t), H_N(t)]\ne 0$ for $t\,\in\,]0,t_F[$ the Trotterization process will not be ideal. In the end, we want the Trotterized evolution to approximate quantum annealing. Using the adiabatic approximation at step $k$ ($\int_{t_k}^{t_{k}+\Delta t_k} H(t) dt \simeq H(t_k)\Delta t_k$), we can write the evolution in this single step as

\begin{equation}
    U_N^{(k)}\ U_L^{(k)} = \text{exp}\big\{-i\, H_N(t_k)\, \Delta t_k\big\}\ \text{exp} \big\{ -i\, H_L(t_k)\, \Delta t_k\big\}.
\end{equation}
This evolution can be considered continuous if there exists an equivalent Hamiltonian $H_{eq}(t)$. It corresponds to the Hamiltonian that controls the evolution as if it was a continuous process governed by the Schr\"odinger equation. We want this equivalent Hamiltonian to converge to the annealing Hamiltonian $H(t)$ as the step size decreases. 
The Baker-Campbell-Hausdorff expansion allows us to write an expression for $H_{eq}(t_k)$\cite{thompsonConvergenceProofGoldberg1989} . 
\begin{equation}
    \begin{split}
        -i\,H_{eq}(t_k)\, \Delta t_k &= \text{log} \Big[\text{exp}  \big\{-i\, H_N(t_k)\, \Delta t_k\big\}\,\\
        & \qquad\qquad \cdot \, \text{exp} \big\{ -i\, H_L(t_k)\, \Delta t_k\big\} \Big]\\
        & = -i\, H(t_k)\, \Delta t_k + \sum_{n\geq 2} \frac{1}{n} \sum_{|\omega|=n} g_\omega\ [\omega]
    \end{split} \label{eq:supp:equiv-hamiltonian}
\end{equation}
where $\omega=\omega_0\omega_1...\omega_{n-1}$ is a word of length $n=|\omega|$ and $\omega_i$ can be $X$ or $Y$, where
\begin{align*}
    X &= -i\ H_L(t_k)\ \Delta t_k, \\
    Y &= -i\ H_N(t_k)\ \Delta t_k.
\end{align*}
$[\omega] = \left[[\omega_0, \omega_1], ... \omega_{n-1}\right]$ represents nested commutators of the word $\omega$ and $g_\omega$ is a scalar coeficient which depends on $\omega$. We want to study the convergence properties of $||H_{eq}(t)-H(t)||$. With this aim, we use the relation \cref{eq:methods:convergence-proof} obtained by reference \cite{thompsonConvergenceProofGoldberg1989}. It makes use of the Lie norm,  satisfying \cref{eq:apx:LieNorm}. 
\begin{equation}
    \left|\left|\, [X, Y]\, \right|\right|_L \leq || X||_L\ ||Y||_L \label{eq:apx:LieNorm}
\end{equation}
\begin{equation}
    \left|\left| \frac{1}{n}\sum_{|\omega|=n} g_\omega \ [\omega]\right|\right|_L\ \leq\ \frac{2 \mathcal{M}^n}{n} \label{eq:methods:convergence-proof}
\end{equation}
where
\begin{equation}
    \begin{split}
        \mathcal{M}&=\max\big(||X||_L, ||Y||_L\big)\\
        &= \Delta t_k\,\cdot \,\max \left( ||H_L(t_k)||_L, ||H_N(t_k)||_L \right)\\
        & = \frac{\Delta t_k}{\Delta t_{M,k}}
    \end{split} 
\end{equation}
Consequently, we obtain a bound of the Hamiltonian difference at step $k$

\begin{equation}
    \begin{split}
        \left|\left| H_{eq}(t_k)\,-\, H(t_k) \right|\right|_L \Delta t_k &\leq \sum_{n\geq 2} \frac{2}{n} \left(\frac{\Delta t_k}{\Delta t_{M,k}}\right)^n\\
    \end{split}  \label{eq:supp:hequiv-series}
\end{equation}
where $H(t)=H_L(t)+H_N(t)$ is the adiabatic Hamiltonian. Two observations can be made about this bound. The first is that for 
\begin{equation}
    \Delta t_k<\Delta t_{M,k}, \label{eq:supp:conv-condition}
\end{equation}
the absolute series converges. The divergence of the absolute series does not guarantee the divergence of the series itself, i.e. the series may conditionally converge for $\Delta t_k\ge\Delta t_{M,k}$. In the case that the series describing $H_{eq}(t)$ diverges, then the evolution cannot be described by the Schr\"odinger equation, and thus is not equivalent to a continuous process. The second observation is that when $\Delta t_k/\Delta t_{M,k} \ll 1$, we may only consider the first term in the series (\ref{eq:supp:hequiv-series}), and get the bound
\begin{equation}
    ||H_{eq}(t_k) - H(t_k)||_L \leq \frac{\Delta t_k}{(\Delta t_{M, k})^2}. \label{eq:supp:hequiv-conv}
\end{equation}    
This indicates that the equivalent Hamiltonian $H_{eq}(t_k)$ gets arbitrarily close to the annealing Hamiltonian $H(t_k)$  for arbitrarily small $\Delta t_k$. For simplicity, we now consider an uniform step size $\Delta t_k = \Delta t$. The convergence condition (\ref{eq:supp:conv-condition}) becomes
\begin{equation}
    \Delta t < \Delta t_M = \min_k t_{M,k}= \min_k \left( ||H_L(t_k)||_L^{-1}, ||H_N(t_k)||_N^{-1} \right) 
\end{equation}
which is the relation shown in the main text.

At this point we have found that the equivalent Hamiltonian converges to the annealing Hamiltonian for each of the steps, but we can also verify this for the whole evolution $0<t<t_F$. 
Thus we now want to compute the norm of the difference between the adiabatic evolution and the equivalent evolution
\begin{equation}
    \begin{split}
        &\left|\left| \mathcal{T}_k \prod_{k=0}^{M-1} \exp\{-i H(t_k)\Delta t\} - \mathcal{T}_k \prod_{k=0}^{M-1} \exp\{-i H_{eq}(t_k)\Delta t\}   \right| \right|\\
        & = \left|\left| \mathcal{T}_k \prod_{k=0}^{M-1} U^{(k)} - \mathcal{T}_k \prod_{k=0}^{M-1} U^{(k)}_{eq}  \right| \right|
    \end{split}
\end{equation}
Firstly, by adding a term that cancels with itself and applying the triangular inequality, we can obtain
\begin{align*}
    \left|\left| \mathcal{T}_k \prod_{k=0}^{M-1} U^{(k)} - \mathcal{T}_k \prod_{k=0}^{M-1} U^{(k)}_{eq}  \right| \right|\leq M \max_k \left|\left| U^{(k)}-U^{(k)}_{eq} \right|\right|.
\end{align*}
If we assume $\Delta t/\Delta t_M\ll 1$ and use \cref{eq:supp:hequiv-conv}, we get
\begin{equation}
    \begin{split}
        \left| \left| U^{(k)} - U^{(k)}_{eq} \right|\right| \simeq \left|\left| H(t_k) - H_{eq}(t_k) \right|\right| \Delta t \leq \frac{\Delta t^2}{(\Delta t_{M})^2}
    \end{split} \label{eq:methods:Ustep-diff}
\end{equation}
Joining the past two relations, we have 
\begin{equation}
    \left|\left| \mathcal{T}_k \prod_{k=0}^{M-1} U^{(k)} - \mathcal{T}_k \prod_{k=0}^{M-1} U^{(k)}_{eq}  \right| \right| \leq t_F\, \frac{\Delta t}{(\Delta t_M)^2} \label{eq:methods:evol_diff}
\end{equation}
This expression assures us that with small enough $\Delta t$ , the whole Trotterized adiabatic evolution will arbitrarily approximate the adiabatic evolution. With noise, we would expect an additional term appear in the expression of $H_{eq}(t_k) \Delta t_k$ (in \cref{eq:supp:equiv-hamiltonian}) that would not depend on $\Delta t$, ultimately implying that in \cref{eq:methods:evol_diff} there would be a constant term independent of $\Delta t$. This suggests that the noisy Trotterized evolution does not approximate the ideal adiabatic process.

\subsection{Probability of measuring the ideal state} \label{apxsect:noise-derivations}

\begin{figure}[t]
    \centering
    \includegraphics[width=\linewidth]{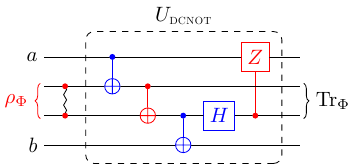}
    \caption{CNOT telegate circuit, equivalent to the one shown in \cref{fig:DCNOT-circuit}. Here, the measurement operations have been swapped with control such that measurement occurs at the end of the circuit. The measurement operation over the entanglement qubits is equivalent to taking the partial trace Tr$_\Phi$. }
    \label{fig:apx:DCNOT-equiv}
\end{figure}

    

A generic distributed operation consists of a sequence of local unitaries and CNOT telegates. The noise is considered to come solely from the shared quantum state 
\[
    \rho_\Phi = x \ketbra*{\Phi}{\Phi}\ +\ \frac{1-x}{4} \mathbb{I}_\Phi,
\]
where $\ket{\phi} = (\ket{00}+\ket{11})/\sqrt{2}$ for the CNOT telegate and $$x=\frac{4}{3}F_\Phi - \frac{1}{3},$$ with $F_\Phi$ being the fidelity between $\rho_\Phi$ and $\ketbra{\Phi}{\Phi}$. 
The CNOT distribution protocol is shown in figs. \ref{fig:DCNOT-circuit}\, and \ref{fig:apx:DCNOT-equiv}, and can be written as the operation
\begin{equation}
    \begin{split}
        \text{DCNOT}[\rho] &\equiv \text{Tr}_{\Phi} \left[ U_{\text{DCNOT}} (\rho \otimes \rho_\Phi) U_{\text{DCNOT}}^\dagger \right]\\
        & = x\ \text{CNOT}[\rho] + (1-x)\ \Theta [\rho]
    \end{split} \label{eq:apx:DCNOT-operation}
\end{equation}
with 
\[
    \Theta [\rho] \equiv \frac{1}{4} \text{Tr}_\Phi \left[ U_{\text{DCNOT}} (\rho \otimes \mathbb{I}_\Phi) U_{\text{DCNOT}}^\dagger \right].
\]
The CNOT$[\rho]$ operation corresponds to the state after ideal distribution, without noise, while $\Theta[\rho]$ is the state after a distribution fault occurs, i.e. with $\rho_\Phi=\mathbb{I}_\Phi/4$. 

Consider $\mathbb{D}_k[\rho]$ the state that results from the distribution of the initial state $\rho$, being the average of all possible $k$ distribution faults in $D$ distribution steps. Note that Tr $\mathbb{D}_k[\rho] = 1$, since $\mathbb{D}_k[\rho]$  is a state. A state $\rho_2$ that has to use $2$ noisy DCNOT operations  can be rewriten as
\begin{equation}
    \rho_2 = x^2\ \mathbb{D}_0[\rho_0]\ +\ 2\,x\,(1-x)\ \mathbb{D}_1[\rho_0]\ +\ (1-x)^2\ \mathbb{D}_2 [\rho_0]. 
\end{equation}
\begin{align*}
    \mathbb{D}_0[\rho_0] &= U_2 \circ CNOT \circ U_1 \circ CNOT \circ U_0 [\rho_0]\\
    \begin{split}
        \mathbb{D}_1[\rho_0] & = \frac{1}{2}\,U_2 \circ \Theta \circ U_1 \circ CNOT \circ U_0 [\rho_0]\\  &+ \frac{1}{2}\,U_2 \circ CNOT \circ U_1 \circ \Theta \circ U_0 [\rho_0]
    \end{split}\\
    \mathbb{D}_2[\rho_0] &= U_2 \circ \Theta \circ U_1 \circ \Theta \circ U_0 [\rho_0]
\end{align*}
with $U_i[\rho]\equiv U_i \rho U_i^\dagger$ being an arbitrary 2-qubit unitary gate operation (noiseless) and $f\circ g [x] \equiv f[g[x]]$ is the composition operation. The probability of a perfect distributed operation (telegate) is given by $x$, and so the state $\rho_0$ after $D$ distribution steps is given by the binomial distribution:

\begin{equation}
    \rho_{D} = \sum^{D}_{k=0} \begin{pmatrix}
        D\\
        k
    \end{pmatrix}\ x^{D-k}\ (1-x)^k\ \mathbb{D}_k[\rho_0] \label{eq:methods:distribution-binomial} 
\end{equation}
If we assume that sucessive faults reduce the probability of measuring the ideal state $\ketbra*{\Psi}{\Psi}=\mathbb{D}_0[\rho_0]$ by $\beta$, we have that $\bra{\Psi} \mathbb{D}_k[\rho_0] \ket{\Psi}=\beta^k$, and so $p_N(F_\Phi)$ becomes
\begin{equation}
    p_N(F_\Phi) = \left(1-\frac{4}{3}(1-F_\Phi)(1-\beta)\right)^D
\end{equation}
This $\beta$ parameter acts as an indicator of the distribution hardness, and in the worst case scenario $\beta=1$ and so we get the lower bound
\begin{equation}
    p_N(F_\Phi)\ > x^{D} = \left(1-\frac{4}{3}(1-F_\Phi)\right)^{D} \label{eq:methods:distribution-probability-no-faults}
\end{equation}

To understand the meaning of $\beta$, we can compute it with the following expression
\begin{equation}
    \beta = \bra{\Psi} \mathbb{D}_1[\rho_0] \ket{\Psi}
\end{equation}
which corresponds to the probability of measuring the ideal state having one distribution fault. In the case that the process corresponds to a single DCNOT operation, then 
\begin{equation}
    \beta = \bra{\Psi} \Theta [\rho_0] \ket{\Psi} \label{apx:noise-beta-overlap}
\end{equation}
Consider the following
\begin{align}
    \rho_0 &= \rho_A \otimes \rho_B\label{apx:noise-rho0}\\
    \rho_{A,B} &= \ketbra*{\Psi_{A, B}}{\Psi_{A,B}}\label{apx:noise-rhoAB}\\
    \ket{\Psi_A} &= a \ket{0}_A + b \ket{1}_A\label{apx:noise-psiA}\\
    \ket{\Psi_B} &= \cos \theta/2 \ket{0}_B + e^{i\phi} \sin \theta/2 \ket{1}_B\label{apx:noise-psiB}\\ 
    \ket{\Psi} &= CNOT[\ket{\Psi_A}, \ket{\Psi_B}] \label{apx:noise-cnot}
\end{align}
where $CNOT[\ket{\Psi_A}, \ket{\Psi_B}]$ represents the CNOT operation with $\ket{\Psi_A}$ as control and $\ket{\Psi_B}$ as target. Using \cref{apx:noise-beta-overlap} and \cref{apx:noise-rho0,apx:noise-rhoAB,apx:noise-psiA,apx:noise-psiB,apx:noise-cnot} we get the following expression
\begin{equation}
    \beta = 1 - 2|a|^2(1-|a|^2) (1-\sin \theta \cos \phi).
\end{equation}
From this we can see that for $\beta=1$, then $\ket{\Psi_A}=\ket{0}_A$ or $\ket{1}_A$ or $\ket{\Psi_B}=\ket{\pm}_B$. This corresponds to a separable operation , no entanglement is generated in either case. For the worst case of $\beta=0$, $\ket{\Psi_A}=\ket{\pm}_A$ and either $\ket{\Psi_B}=\ket{0}_B, \ket{1}_B$ or $\ket{\Psi_B}= (\ket{0}_B \pm i \ket{1}_B)/\sqrt{2}=\ket{\pm y}_B$. This results in an operation that generates non-separable states, such as the Bell states. It is intuitive that a fault in the distribution cannot generate entangled states.

In essence, $\beta$ is a measure of the importance of the distributed operations relative to the whole process. If the distributed operations do not generate entangled states, then $\beta$ will be close to 1, otherwise it will near 0. Another way to interpret it, under the process of distributed quantum annealing, is that $\beta$ is a way to measure the relevance of the non-local couplings to the dynamics of the annealing system.


\section{Numerical simulations} \label{apxsect:numerical}

\begin{figure}
    \centering
    \includegraphics[width=0.8\linewidth]{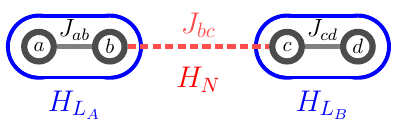}
    \caption{ Coupling graph of the 4 qubit spin chain model, with labels $\Omega=\{a,b,c,d\}$. The local Hamiltonian $H_L=H_{L_A}+H_{L_B}$ has couplings $J_{ab}\sigma^z_a\sigma^z_b$ and $J_{cd}\sigma^z_c\sigma^z_d$, and the non-local Hamiltonian $H_N$ only contains $J_{bc}\sigma^z_b\sigma^z_c$. Although being simple, it allows us to explore the behaviour of the protocol.} \label{fig:four-qubit-graph}
\end{figure}

We now look more quantitatively at how the method performs on a four qubit spin chain. The corresponding representation is shown in \cref{fig:four-qubit-graph}. Its qubits are labeled $\Omega=\{a, b, c, d\}$, and so the local and non-local Hamiltonians are

\begin{align}
    H_L(t) &= \left(1-\frac{t}{t_F}\right) \sum_{i\in\Omega} \sigma_i^x\ +\ \frac{t}{t_F} \sum_{(i,j)\in e_L} J_{ij}\sigma^z_i\sigma^z_j  \\
    H_N(t) &= \frac{t}{t_F}\ J_{bc} \sigma^z_b \sigma^z_c
\end{align}
with $e_L=\{(a,b), (c,d)\}$, $J_{ab}=J_{cd}=1$ and $J_{bc}=-2$.

We use the relative energy error $\varepsilon_{D0}$ (\ref{eq:rel-en-error}) to quantify the perfomance of the method. It corresponds to the energy difference between $E_D=\bra{\psi_D} H_F\ket{\psi_D}$ (energy of the state reached by distributed protocol) and $E_0$ (the ground state energy of $H_F$). It is normalized by $E_{\text{mix}}-E_0$, where $E_{\text{mix}}=\bra{\phi_0}H_F\ket{\phi_0}$ is the energy of the inital state $\phi_0$ measured in $H_F$. $E_{\text{mix}}$ is the average energy of a randomly choosen state, taking in consideration the inital Hamiltonian. 

\begin{equation}
    \varepsilon_{D0} = \frac{E_D-E_0}{E_{\text{mix}} - E_0 } \label{eq:rel-en-error}
\end{equation}

With this definition of the normalized energy error, we have that $\varepsilon_{D0}\sim0$ when the evolution successfully finds the ground state (or one of the ground states, if $H_F$ is degenerate), and $\varepsilon_{D0}\sim1$ when no useful information about the ground state can be obtained from $\ket{\psi_{D}}$.  

\begin{figure}
    \includegraphics[width=\linewidth]{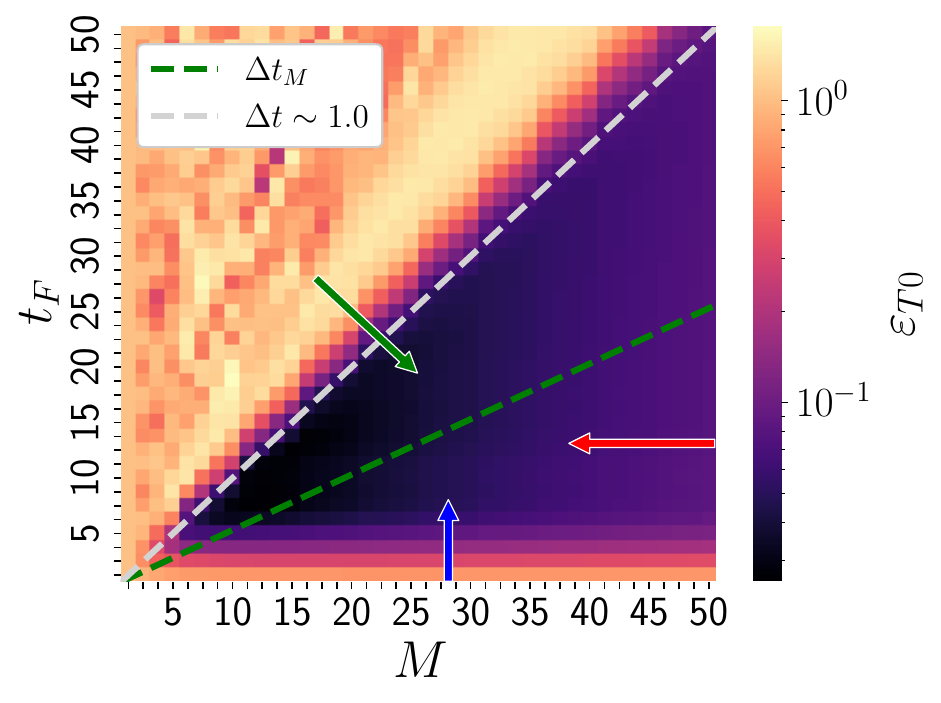}
    \caption{ Relative energy error of the Trotter process, varying the annealing time $t_F$, the number of Trotter steps $M$, and considering noisy distribution with $F_\Phi=0.999$. The local fields are set $h_i=0 \ \forall i\in\Omega$ and couplings $J_{ab}=J_{cd}=1$, $J_{bc}=-2$. The three arrows disclose the effect of the tunnable parameters in the Trotterization process. The blue arrow points towards a gradient of the error due to improved adiabaticity conditions, as $t_F$ grows. The green arrow points across the phase transition exhibited by this process. The gradient directed by the red arrow is a result of noisy distribution: as the number of distribution steps increases, also do the errors related to distribution. The theoretical convergence $\Delta t_M$ is drawn with a dashed green line and an estimation for the transition is drawn with a light gray dashed line. \label{fig:spin-chain-heatmap}\label{fig:results:4toy1} }
\end{figure}

There are 3 tunnable parameters related to the Trotterization process: the annealing time $\boldsymbol{t_F}$, which controls the speed of the adiabatic evolution; the number of Trotter steps $\boldsymbol{M}$, or equivalently the Trotter step size $\boldsymbol{\Delta t}=t_F/M$, which control the accuracy of the Trotterization; and the noise of the distribution process, tuned with the entanglement fidelity $\boldsymbol{F_\Phi}$. The overall effect of $t_F$ and $M$ in a noisy scenario ($F_\Phi=0.999$) can be seen in \cref{fig:spin-chain-heatmap}, indicated with different colored arrows. Firstly, there is a clear transition region at constant $\Delta t$, which was expected from the theoretical analysis. This transition is visible at roughly $\Delta t\sim 1.0$, while the theoretical prediction is 
\[
    \begin{split}
        \Delta t_M&\equiv\min_k\Big(\big|\big|H_L(t_k)\big|\big|_L^{-1}, \big|\big|H_N(t_k)\big|\big|_L^{-1}\Big) = 0.5,
    \end{split}    
\] 
being less than the one observed in the simulated Trotterization. The region where the algorithm appears to converge for $\Delta t>\Delta t_M$ is likely related to the conditional convergence of the equivalent Hamiltonian $H_{eq}$, where our theoretical analysis cannot establish any bound. Secondly, a gradient is seen as the annealing time $t_F$ increases and the relative energy error $\varepsilon_{D0}$ decreases. This effect is due to the adiabatic nature of the process: the slower the process is, the closer the final state is to the ground state of $H_F$. Thirdly, a gradient is visible as the number of Trotter steps $M$ increases. This is due to distribution noise. The probability of measuring the anneal state $p_N(F_\Phi)$ (seen from the lower bound \cref{eq:anneal-measure-probability}) decreases with $M$, and so the distributed state energy $E_D$ increases above the anneal state energy, and more importantly above $E_0$, leading to the increase of $\varepsilon_{D0}$. 

\begin{figure}
    \centering
    \includegraphics[width=0.9\linewidth]{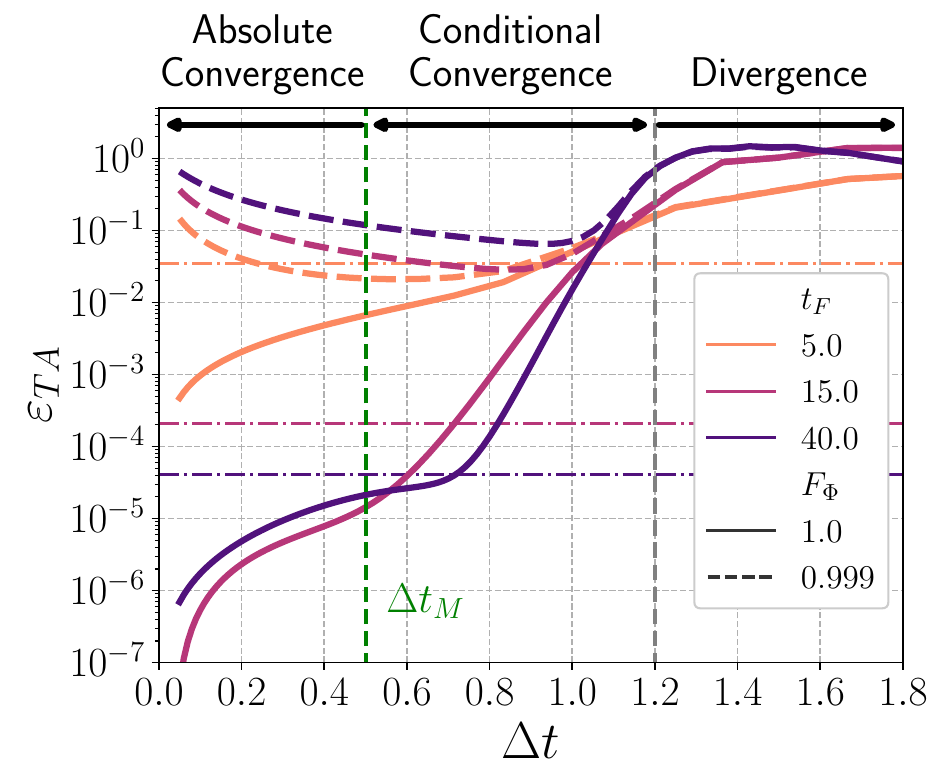}
    \caption{Relative energy error between the Trotterized process and the ideal annealing evolution, with fixed annealing times indicated in different colors. The horizontal dash-dotted lines marks the annealing error $\varepsilon_{A0}$, for each annealing time $t_F$. At the convergence transition point, i.e. $\Delta t=\Delta t_M$, the Trotter energy error $\varepsilon_{TA}<\varepsilon_{A0}$ for ideal distribution, visible by comparing the thick lines with the horizontal dash-dotted lines in the respective color. The annealing error at the transition point is bigger than the Trotter error, meaning that the convergence condition appears to be enough to select an adequate $\Delta t$. }
    \label{fig:results:4toy-2}
\end{figure}

\begin{figure}
    \centering
    \includegraphics[width=0.9\linewidth]{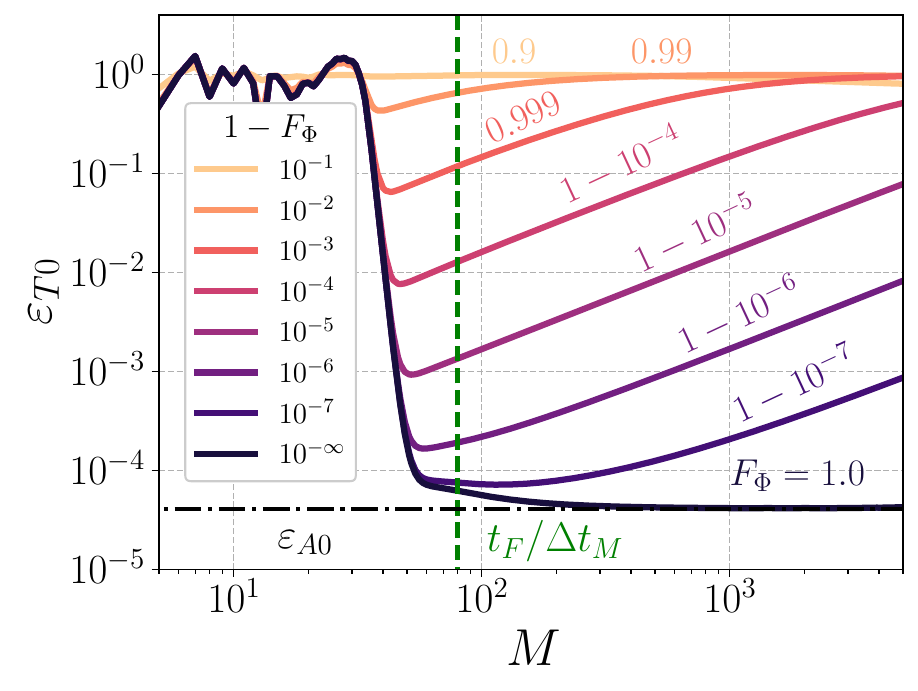}
    \caption{Relative energy error $\varepsilon_{T0}$ for a range of values of $F_\Phi$ and fixed $t_F$. The dashed vertical line marks $M=t_F/\Delta t_M$ associated with the transition, and the horizontal dash-dotted line marks the annealing error $\varepsilon_{A0}$. With ideal distribution, i.e. $F_\Phi=1$, the distributed quantum annealing process approximates quantum annealing arbitrily well, with larger $M$. This is not true when there is distribution noise $F_\Phi<1$. In this case, the Trotter energy error does not get as low as the annealing energy error, and gets further and further away as the noise increases (with $F_\Phi$ decreasing).}
    \label{fig:results:4toy-3}
\end{figure}


Due to the adiabaticity condition (slow evolution, large $t_F$) the Trotterized adiabatic evolution is similar to the evolution of Floquet systems for intermediate time intervals $t\in [\tau, \tau+\Delta \tau]$, $\tau\in[0, t_F-\Delta \tau]$ with $\Delta t \ll \Delta \tau\ll t_F$. Indeed, the phase transition we have observed is also a common feature of Floquet systems \cite{dalessioManybodyEnergyLocalization2013,kuwaharaFloquetMagnusTheoryGeneric2016,heylQuantumLocalizationBounds2019,thanasilpQuantumSupremacyQuantum2021}. In the literature this phase transition is known as the many-body localization (MBL) phase transition. The region that remains at low energy, i.e. where $\varepsilon_{D0}\sim0$, is analogous to the localized phase in the MBL transition, and is associated with low ergodicity. In contrast, the region where $\varepsilon_{D0}\sim 1$ has energy close to the average of the eigenenergies of the Hamiltonian, and so is analogous to a system with infinite temperature, i.e. high ergodicity. Thus it is commonly refered to as the thermalization phase. In Floquet systems, this transition is due to the divergence of the Floquet Hamiltonian, which is constructed in a similar way to the equivalent Hamiltonian from \cref{eq:equiv-Hamiltonian}. 

\section{Secure multi-party annealing - Linear Least Squares (LLSq)} \label{apx:sect:linear-least-squares}

Linear least squares is a typical way of solving over defined systems of linear equations, which appear often in logistics, finances and scientific data analysis. When expressed in QUBO form, these can be solved by quantum annealers, although in general they may require arbitrary qubit connectivity \cite{changLeastsquaresSolutionsPolynomial2019,BorleHowProblems}. This requirement can make the problem hard to embed onto the quantum annealing hardware. The addition of non-local couplings can potentially alleviate this issue in different ways, by adding either non-local energy or duplication couplings.

The QUBO form of the LLSq algorithm is generated as follows. Consider $A \vec x = \vec b$, where $A\in \mathbb{R}^{n\times m}$ is a matrix where $n\ge m$, $\vec b \in \mathbb{R}^n$ a vector and $x\in \mathbb{R}^m$ is the unknown vector to be determined. The aim of LLSq is to find the minimum of $|| A \vec x - \vec b||^2$ over all possible $\vec x$. Consider $[A]_{ij} = A_{ij}$ the row, column elements of matrix $A$ and $[\vec x]_j=x_j$,  $[ \vec b]_i=b_i$ the elements of the vectors $\vec x, \vec b$ respectively. The quantity to be minimized is $E=|| A \vec x - \vec b||^2$ and can be written as
\begin{equation}
    \begin{split}
        E &= \sum_j \left[\left(\sum_i A_{ij}^2\right) - 2 \sum_i b_i A_{ij}\right] x_j \\
        &+ \sum_{k<l} \left[2 \sum_i A_{ik} A_{il}\right] x_k x_j
    \end{split}
\end{equation}
which is almost in QUBO form. The real values $x_j$ has to be mapped to the binary representations $q_{jl}$
\begin{equation}
    x_j \rightarrow \sum_l c_l q_{jl}
\end{equation}
where $c_l$ is the encoding factor. For example, if the problem was defined such that $0<x_j<1$ we could choose $c_l = 2^{-l}, l\ge 0$ \cite{OMalley2016ToQ.jl:Julia}. In general, better solution precision requires additional qubits (for each $x_j$). Then, this QUBO problem is mapped to the Ising model by doing $q \rightarrow (1-\sigma^z)/2$. 

In a two-party LLSq problem there are two distrusting agents with different constraints represented by the matrices $A^{(i)}$ and vectors $\vec b^{(i)}$, with $i=1,2$ distinguishing the constraints for each agent. Their aim is to solve the linear equations $A^{(i)} \vec x^{(i)} = b^{(i)}$, with the additional objective that they share the same solution $\vec x^{(1)}=\vec x^{(2)}$, while not revealing secret information contained in their constraints $A^{(i)}, \vec b^{(i)}$. This kind of problem is often seen in distributed databases with secure data, and there are a few classical ways of tackling the problem \cite{Karr2007SecureDatabases,Qiu2020Privacy-PreservingMasking}. Here we aim to show that our proposal for a quantum annealing system enhanced with non-local operations is naturally capable of tackling this optimization problem with the prospect of the potential quantum speedup brought by quantum annealing. In matrix form, this optimization is expressed as the following

\begin{equation}
    \begin{bmatrix}
        A^{(1)}\\
        A^{(2)}
    \end{bmatrix} \vec x = \begin{bmatrix}
        \vec b^{(1)}\\
        \vec b^{(2)}
    \end{bmatrix}
\end{equation}
with the aim of finding the unknown $\vec x$. In LLSq, this corresponds to minimizing the energy
\begin{equation}
\begin{split}
    E_{1,2} =&\left|\left|\begin{matrix}
        A^{(1)} \vec x - b^{(1)} \\
        A^{(2)} \vec x - b^{(2)} 
    \end{matrix}\right|\right|^2  \\
         =&\, || A^{(1)} \vec x - b^{(1)} ||^2 + || A^{(2)} \vec x - b^{(2)} ||^2\\
         +& \ 2 (A^{(1)} \vec x - b^{(1)} )\cdot(A^{(2)} \vec x - b^{(2)} ) 
    \end{split}
\end{equation}
 
The final energy term contains constraints from both parties, but is not necessary to encode the ground state representing the solution that is being searched. In essence, this term gives the 'search algorithm' a preferential direction for the optimization.
Removing the final constraint, we can have distinct quantum annealers running each of the optimization problems for each of the parties. The solution $\vec x$ is assured to be the same if the qubits are duplicated, as described in \ref{sec:duplication-coupling}. This two party process can also be extended to $N$ distrusting parties through qubit $n$-plication. 

\bibliographystyle{unsrtnat}
\bibliography{library,references}

\end{document}